# Physics-informed EDFA Gain Model Based on Active Learning

Xiaomin Liu[(1)], Yuli Chen[(1)], Yihao Zhang[(1)], Yichen Liu[(1)], Lilin Yi[(1)], Weisheng Hu[(1)], Qunbi Zhuge[(1)]

(1) Shanghai Jiao Tong University, qunbi.zhuge@sjtu.edu.cn

**Abstract** *We propose a physics-informed EDFA gain model based on the active learning method. Experimental results show that the proposed modelling method can reach a higher optimal accuracy and reduce ~90% training data to achieve the same performance compared with the conventional method.*

**Introduction**

Erbium-doped fiber amplifier (EDFA) is an important device for modern optical communication systems. Building an accurate digital model for an EDFA can efficiently assist the physical layer control and management [1]. However, due to the complex structure and manufacturing discrepancy, the theoretical EDFA model cannot reach a high accuracy. Fortunately, the artificial intelligence (AI) outperforms the analytical model to achieve better performance in many situations [2], [3]. However, the AI-based methods are purely data-driven, which means they are highly dependent on the quantity and quality of the training data. For precise modelling, the AI-based methods require a large amount of measured gain spectra for training the model. In [3], the model based on the neural network (NN) utilizes about 50000 training data. In [4], the training data size is about 25000. The high requirement on the data size hinders the practical deployment of the AI-based modeling methods. First, it is hard to prepare a large dataset for each EDFA for the customized modeling. Second, the large dataset increases the training burden of the AI-model, and the inaccuracy during the data collection can negatively influence the modeling performance.

As a result, the aim of building the AI-based EDFA model is to make the accuracy as high as possible with a training dataset as small as possible [5]. To address the abovementioned problems, in this paper, we propose a physics-informed EDFA gain model based on the active learning (AL) sampling strategy for efficiently selecting the most useful gain spectra for training. An automatic spectra measurement system is built to demonstrate the performance of the proposed modeling methods experimentally. The experimental results indicate that, using about 100 measured spectra, the proposed method can build an EDFA gain model with the accuracy higher than the traditional NN-based method. Modeling performance under different sampling numbers per training iteration are also investigated in this paper to show the stability of the proposed modeling method.

**The Architecture of the EDFA Measurement Strategy Based on Active Learning**

Fig.1(a) shows the architecture of the spectra measurement strategy based on the active learning. First, a dataset containing a small number of the gain spectra is collected by the automatic EDFA measuring system. Based on the pre-collected dataset, a pre-trained model can be initialized. Meanwhile, the repository of the candidate to-be-measured input spectra is generated randomly. Afterwards, the pre-trained EDFA gain model is employed to make estimations on the repository of the candidate input spectra. The pre-trained model can make estimations and provide the corresponding confidence degree. The to-be-measured spectrum with the lowest confidence degree is chosen as the next spectrum for measuring. The measured gain spectra are added to the pre-training dataset and the model is updated. Iteratively, the training dataset is enlarged and the model can be updated to achieve a higher accuracy.

**The Physics-informed Model Construction and AL-based Acquisition Function**

Active learning has been approved to be effective for modeling the optical system [5]. To realize the active learning strategy, we use the uncertainty-based data acquisition method introduced in the previous section. The utilized machine learning method should be capable of: i) making estimations and providing the uncertainty degree of the estimations, and ii) integrating the analytical model as the prior to fuse the physics-based model. To achieve these goals, we use the Gaussian process regression (GPR) to build the data-driven EDFA model. The GPR is a non-parametric model and can employ the analytical model as the priors [6]. Additionally, it can provide the mean and variance of the estimation, which can be regarded as the predicted value and the uncertainty degree, respectively. If the input features are represented as the $n$-dimensional

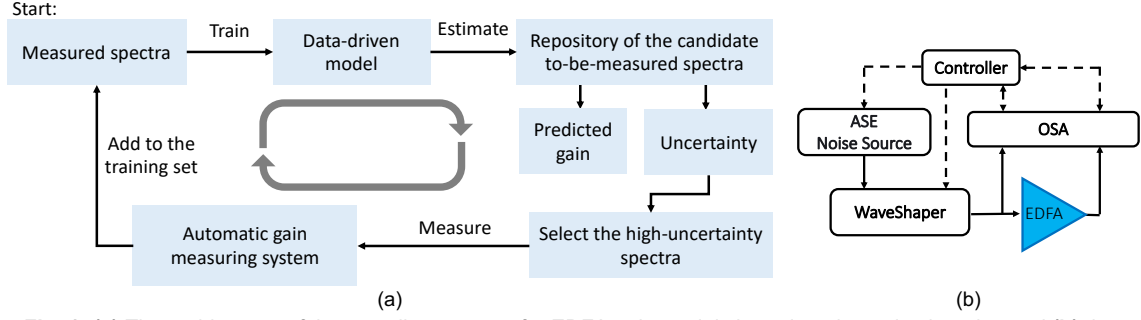

**Fig. 1: (a)** The architecture of the sampling strategy for EDFA gain models based on the active learning and **(b)** the experiment setup for sampling the gain spectra.

matrix of $X$, they can be represented as

$$X = \begin{pmatrix} x_{11} & x_{12} & \cdots & x_{1n} \\ x_{21} & x_{22} & \cdots & x_{2n} \\ \vdots & \vdots & \vdots & \vdots \\ x_{m1} & x_{m2} & \cdots & x_{mn} \end{pmatrix}, \quad (1)$$

where the $x_{ij}, i = 1\ldots m, j = 1\ldots n$, represents the $i$-th feature of the $j$-th sample. The matrix of estimated gain spectra is $Y$, which can be written as

$$Y = \begin{pmatrix} y_{11} & y_{12} & \cdots & y_{1n} \\ y_{21} & y_{22} & \cdots & y_{2n} \\ \vdots & \vdots & \vdots & \vdots \\ y_{m1} & y_{m2} & \cdots & y_{mn} \end{pmatrix}, \quad (2)$$

where the $y_{ij}, i = 1\ldots m, j = 1\ldots n$ is the estimated $i$-th output of the $j$-th sample. In our work, each $x$ and $y$ represent the per-channel input spectra and corresponding gain spectra, respectively. The mapping between them, denoted as $f(x)$, can be seemed as the Gaussian process (GP) of

$$f(x) \sim GP(m(x), k(x,x)), \quad (3)$$

where the $m(x)$ is the mean function of the model and $k(x,x)$ is the covariance function evaluating the similarity of each data. For the to-be-estimated input spectrum, which can be written as $x_*$, its estimation $f(x_*)$ follows the joint Gaussian process of

$$\begin{bmatrix} Y \\ f(x_*) \end{bmatrix} \sim N(\begin{bmatrix} \mu(x) \\ \mu(x_*) \end{bmatrix}, \begin{bmatrix} K + \sigma^2 I & k(x, x_*) \\ k(x_*, x) & k(x_*, x_*) \end{bmatrix}), \quad (4)$$

where the $\mu(x)$ and $\mu(x_*)$ are the mean value of the estimation of the spectrum $x$ and $x_*$, respectively. $K$ is the covariance matrix of the training dataset. Therefore, the estimated mean $\mu_*$ and variance $\sigma_*^2$ of $x_*$ are

$$\mu_* = m(x_*) + k(x_*, x)(K + \sigma^2 I)^{-1}(Y - m(x)), \quad (5)$$

$$\sigma_*^2 = k(x_*, x_*) - k(x_*, x)(K + \sigma^2 I)^{-1} k(x, x_*). \quad (6)$$

The GPR is widely used for estimating the performance of optical systems [7]. The strategy for setting the probe channels efficiently based on the GPR is also proposed for fiber modeling [8]. Based on the sampling strategy and the GPR training algorithm, the data collection for the EDFA gain spectra can be conducted efficiently and the EDFA gain model is expected to use these useful and informative data to learn the most precise model. To further improve the modeling performance of the GPR, the mean function $m(x)$ is set as the analytical model of the EDFA in [9], which can be written as

$$m(x_i) = g(x_i) + \frac{\sum_{i=1}^{z}\{G_T - g(x_i)\}}{z}, \quad (7)$$

where the $g(x_i)$ is the gain of the $i$-th channel when all channels are occupied. The $G_T$ is the target gain spectrum containing the whole $z$ channels.

**The Experimental Setup and Model Construction**

We evaluate the performance of the proposed modeling method in an experiment. As shown in Fig. 1(b), an automatic EDFA measuring system is constructed. An amplified spontaneous emission (ASE) noise source is used for generating the flat spectrum. After filtered by the Finisar WaveShaper 4000A programmable optical filter, the in-service signals are established. 80 channels with the spacing of 50 GHz in the C-band are set. Among them, 40 odd-number channels are selected as the signals for measurement while other channels are filtered out. Two optical spectrum analyzers (OSAs) are used for measuring the input and output power spectrum of the EDFA. The operating mode of the measured EDFA is set as the automatic gain control (AGC) mode with a gain of 16 dB.

To generate the repository of to-be-measured spectra, the 40 channels are assumed to be occupied or idle randomly. For the occupied channel, a random deviation of each signal power from -2 dB to 2 dB with a step size of 1 dB is considered by setting the attenuation of the WaveShaper. In total, 9578 to-be-measured input spectra are generated in the repository, which can be deployed as the WaveShaper configuration and then measured by the automatic spectra measurement system. Besides, 1002 heterogeneous signal spectra are generated randomly with the same rule and measured by the automatic experimental system as the testing dataset.

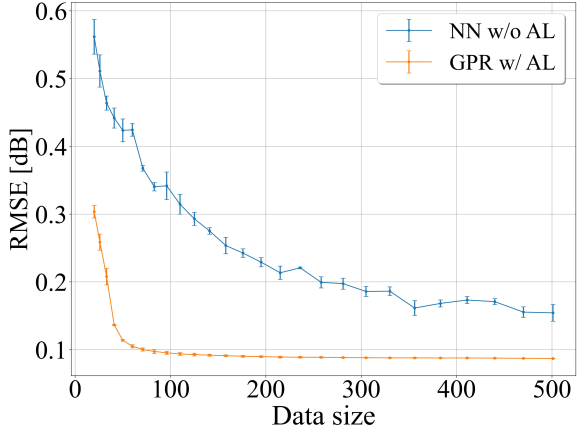

**Fig. 2:** The RMSEs of models with different modeling methods using different sizes of data.

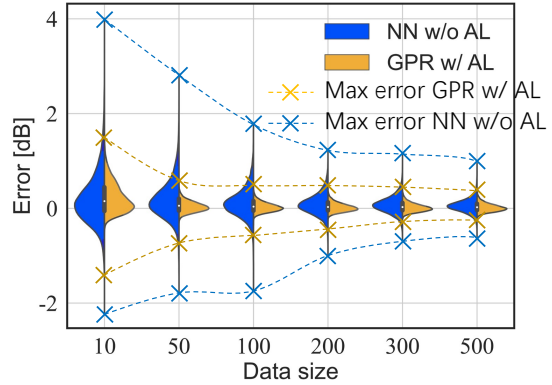

**Fig. 3:** The error histograms and maximum errors of models with different sampling strategies and ML algorithms.

We build the data-driven gain model by the proposed method based on Python Scikit-learn. Firstly, one sample from the candidate repository is measured randomly as the pre-collected training dataset. During the active learning, for each iteration, the number of the measured spectra selected from the repository for training is 1. For comparison, similar to [3], a NN-based traditionally-trained model is constructed with three hidden layers of 128, 64, 32 nodes. The input features of these data-driven models are the same, of which are the signal power of each channel. The output is the estimated gain value of each channel. For each model, the min-max normalization is conducted before training. To reduce the randomness, for each case, we conduct three round experiments and their fluctuations of the corresponding root-mean-square errors (RMSE) are shown by the error bars. What should be mentioned is that since some of the channels are idle, these channels are not considered when calculating the RMSE.

**Results and Discussions**

To investigate the accuracy performance of the proposed method, the RMSEs of the proposed model and traditional model under different sizes of training data are shown in Fig. 2. For the traditionally-trained NN-based models, the estimation errors are large when the data size is small. The best accuracy achieved by 500 samples is about 0.15 dB. On the contrary, the AL-based physics-informed model converges at a high speed and can reduce the RMSE to less than 0.1 dB with about 100 samples, demonstrating its significant learning ability. Moreover, the performance of the proposed model is relatively stable. To achieve the same RMSE on the same testing dataset, the proposed method can largely reduce the training data size, making it possible to prepare a customized tiny dataset for building the precise gain model for each EDFA. The error histograms of the model trained by different methods with different sizes of data are plotted in Fig. 3. The results show that, the model trained by the proposed method can have lower estimation errors and can converge faster, demonstrating the learning ability of selecting

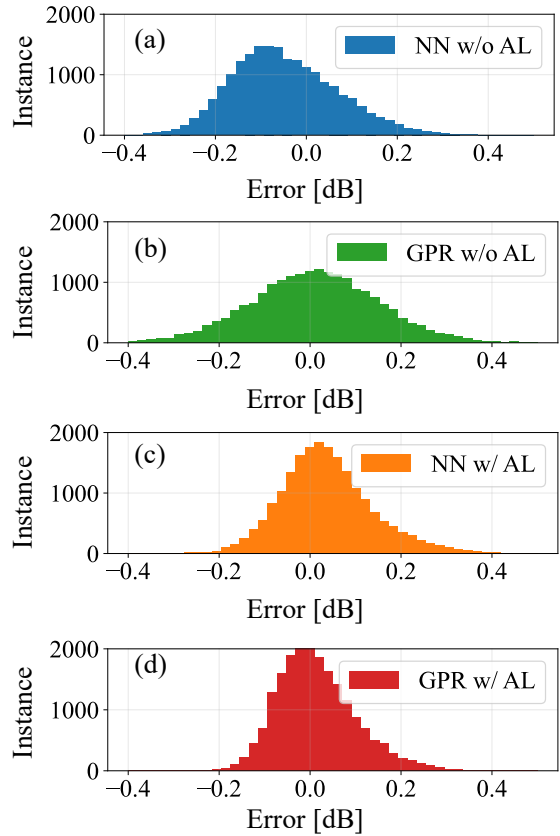

**Fig. 4:** The error histograms of the testing dataset containing 1002 samples estimated by a) the NN without AL, b) the physics-informed GPR without AL, c) the NN with AL-selected data, d) the physics-informed GPR

data and learning.

To evaluate the contributions of the proposed methods for sampling and fusing physical knowledge, we plot the RMSEs of the GPR-based and NN-based models using data selected by active learning or randomly. There are four cases: a) the NN without AL, b) the physics-informed GPR without AL, c) the NN with AL-selected data, d) the physics-informed GPR with AL. The error histograms of each channel trained with different methods on the testing dataset containing 1002 samples are shown in Fig. 4. Results show that compared with NN, the physics-informed GPR-based model can perfectly learn the mean value of the gain spectra. By using the training data selected by AL, both the NN and GPR with AL-selected data can have higher accuracy compared with the model learned from the randomly-selected data. These results demonstrate that the proposed modeling method can improve the performance independently and cooperatively.

**Conclusions**

We propose a physics-informed EDFA gain model based on the active learning. The estimations of the physics-based analytical models are integrated as the prior knowledge for the GPR-based model. The most informative data are selected by the active learning methods to achieve the optimal accuracy while using the least amount of data. In the experiments, compared with the traditional NN-based gain model, the proposed method can efficiently achieve the higher accuracy and significantly reduce the size of data needed for training, making it possible to prepare a precise gain model for each EDFA with few times of measurements.

**References**
[1] D. Rafique and L. Velasco, "Machine learning for network automation: Overview, architecture, and applications [Invited Tutorial]," *J. Opt. Commun. Netw.*, vol. 10, no. 10, pp. D126–D143, 2018, doi: 10.1364/JOCN.10.00D126.
[2] S. Zhu, C. L. Gutterman, W. Mo, Y. Li, G. Zussman, and D. C. Kilper, "Machine Learning Based Prediction of Erbium-Doped Fiber WDM Line Amplifier Gain Spectra," *Eur. Conf. Opt. Commun. ECOC*, vol. 2018-Septe, no. 1, pp. 1–3, 2018, doi: 10.1109/ECOC.2018.8535323.
[3] Y. You, Z. Jiang, and C. Janz, "Machine Learning-Based EDFA Gain Model," *Eur. Conf. Opt. Commun. ECOC*, vol. 2018-Septe, no. 1, pp. 1–3, 2018, doi: 10.1109/ECOC.2018.8535397.
[4] S. Zhu *et al.*, "Hybrid machine learning EDFA model," *Opt. InfoBase Conf. Pap.*, vol. Part F174-, pp. 7–9, 2020, doi: 10.1364/OFC.2020.T4B.4.
[5] A. Vahdat, M. Belbahri, and V. P. Nia, "Active Learning for High-Dimensional Binary Features," *15th Int. Conf. Netw. Serv. Manag. CNSM 2019*, 2019, doi: 10.23919/CNSM46954.2019.9012676.
[6] A. Bartók-Pártay, "Gaussian Process," vol. 384, pp. 23–31, 2010, doi: 10.1007/978-3-642-14067-9_3.
[7] J. W. Nevin, F. J. Vaquero-Caballero, D. J. Ives, and S. J. Savory, "Physics-Informed Gaussian Process Regression for Optical Fiber Communication Systems," *J. Light. Technol.*, vol. 39, no. 21, pp. 6833–6844, 2021, doi: 10.1109/JLT.2021.3106714.
[8] D. Azzimonti, C. Rottondi, and M. Tornatore, "Reducing probes for quality of transmission estimation in optical networks with active learning," *J. Opt. Commun. Netw.*, vol. 12, no. 1, pp. A38–A48, 2020, doi: 10.1364/JOCN.12.000A38.
[9] K. Ishii, J. Kurumida, and S. Namiki, "Experimental investigation of gain offset behavior of feedforward-controlled WDM AGC EDFA under various dynamic wavelength allocations," *IEEE Photonics J.*, vol. 8, no. 1, pp. 1–13, 2016, doi: 10.1109/JPHOT.2016.2514487.